# Olivine-rich exposures at Bellicia and Arruntia craters on (4) Vesta from Dawn FC

Guneshwar Thangjam[1], Andreas Nathues[1], Kurt Mengel[2], Martin Hoffmann[1], Michael Schäfer[1], Vishnu Reddy[1,3], Edward A. Cloutis[4], Ulrich Christensen[1], Holger Sierks[1], Lucille Le Corre[3], Jean-Baptiste Vincent[1], Christopher T. Russell[5]

[1]Max-Planck-Institute for Solar System Research, Justus-von-Liebig-Weg 3, 37077 Göttingen, Germany

[2]Clausthal University of Technology, Adolph-Roemer-Straße 2a, 38678 Clausthal-Zellerfeld, Germany

[3]Planetary Science Institute, 1700 East Fort Lowell, Suite 106, Tucson, AZ 85719-2395, USA

[4]Department of Geography, University of Winnipeg, 515 Portage Avenue Winnipeg, Manitoba, Canada R3B 2E9

[5]Institute of Geophysics and Planetary Physics, University of California, 3845 Slitcher Hall, 603 Charles, USA

Pages: 43
Figures: 12

**Proposed Running Head:** Olivine-rich exposures in Bellicia and Arruntia craters on Vesta

**Editorial correspondence to:**
Guneshwar Thangjam
Justus-von-Liebig-Weg 3,
37077 Göttingen
Germany.
Email: thangjam@mps.mpg.de

**Abstract:**

We present an analysis of the olivine-rich exposures at Bellicia and Arruntia craters using Dawn Framing Camera (FC) color data. Our results confirm the existence of olivine-rich materials at these localities as described by Ammannito et al. (2013a) using Visual Infrared Spectrometer (VIR) data. Analyzing laboratory spectra of various Howardite-Eucrite-Diogenite meteorites, high-Ca pyroxenes, olivines and olivine-orthopyroxene mixtures, we derive three FC spectral band parameters that are indicators of olivine-rich materials. Combining the three band parameters allows us, for the first time, to reliably identify sites showing modal olivine contents >40%. The olivine-rich exposures at Bellicia and Arruntia are mapped using higher spatial resolution FC data. The exposures are located on the slopes of outer/inner crater walls, on the floor of Arruntia, in the ejecta, as well as in nearby fresh small impact craters. The spatial extent of the exposures ranges from a few hundred meters to few kilometers. The olivine-rich exposures are in accordance with both the magma ocean and the serial magmatism model (e.g., Righter and Drake 1997; Yamaguchi et al. 1997). However, it remains unsolved why the olivine-rich materials are mainly concentrated in the northern hemisphere (~36-42° N, 46-74° E) and are almost absent in the Rheasilvia basin.

# 1. Introduction

Asteroid (4) Vesta is geologically the most diverse differentiated chondritic body that remained intact surviving the catastrophic collisional events in the Solar System (e.g., Keil 2002; Russell et al. 2012, 2013). The exploration of such a proto-planetary body enriches the understanding of the geological conditions prevalent in the early Solar System. The observational, meteoritic and dynamical evidences so far suggest that Vesta is the parent body of many of the Howardite-Eucrite-Diogenite (HEDs) meteorites (McCord et al. 1970; Thomas et al. 1997; Migliorini et al. 1997; Schenck et al. 2012; Reddy et al. 2012b; Russell et al. 2012, 2013). Models based on the petrogenesis of HEDs (Ruzicka et al. 1997; Righter and Drake 1997; Warren 1997) favor the evolution of Vesta by an extensive melting (magma ocean). The evolution by sequential development of eruptions in shallow multiple magma chambers (serial magmatism) is also postulated (Yamaguchi et al. 1996, 1997). The serial magmatism on Vesta is consistent with the existing variations of incompatible trace element abundances in diogenites (Mittlefehldt 1994; Fowler et al. 1995; Shearer et al. 1997), and the wide range of Mg- compositions in pyroxene or olivine among olivine-bearing diogenites (Beck and McSween 2010; Shearer et al. 2010). Mandler and Elkins-Tanton (2013) proposed a two-step model of magmatic evolution from a bulk mantle composition based on major and minor elements estimated from earlier studies by Righter and Drake (1997), Dreibus and Wänke (1980), Ruzicka et al. (1997), Lodders (2000), Boesenberg and Delaney (1997). They claimed that their magma ocean model (60-70% equilibrium crystallization followed by fractional crystallization of the residual liquid in shallow magma chambers) can explain the evolution of Vesta in terms of the diverse lithologies/petrogenesis among HEDs. The assumption of an olivine rich mantle of Vesta seems to be justified regardless to the above mentioned models by which various olivine-bearing lithologies like dunites (>90% olivines), harzburgites (40-90% olivines) and olivine-orthopyroxenites (<40% olivine) could be formed in the Vestan mantle or deeper crust (Mandler and Elkins-Tanton, 2013). The exposures of olivine-rich mantle materials were expected in the huge Rheasilvia basin (Thomas et al. 1997; Gaffey 1997; Reddy et al. 2010, 2011a; Beck and McSween 2010; McSween et al. 2011, 2013; Tkalcec et al. 2013; Le Corre et al. 2013). The Rheasilvia basin (~500 km in diameter) superimposes the older Veneneia basin (~400 km in diameter) to a large extent in the southern hemisphere (Schenk et al. 2012).

Recently, olivine has been identified in Bellicia and Arruntia craters by Ammannito et al. (2013a) using Visible and Infrared Spectrometer (VIR) data. VIR is a hyperspectral instrument, which operates in the wavelength range between 0.2 and 5 μm (De Sanctis et al. 2011). The finding of olivine-rich sites in the northern hemisphere (~36-42° N, 46-74° E) despite the absence of such sites in the huge Rheasilvia basin complicates the understanding of the geological evolution of Vesta. Therefore, we mapped and investigated the potential olivine exposures using FC color data to understand the origin and nature of the olivine (Thangjam et al. 2014; Nathues et al. 2014a). The FC instrument (Sierks et al. 2011) acquired color images in 7 filters between 0.44 and 0.96 µm. The spatial resolution of FC exceeds the VIR resolution threefold. Despite FC's limited wavelength range and limited number of filters, the robustness of color parameters in constraining surface composition and mineral identification on Vesta data is well demonstrated (e.g., Le Corre et al. 2011; Reddy et al. 2012b; Thangjam et al. 2013; Nathues et al. 2014b)

*1.1. Olivine in HEDs and pre-Dawn Vesta background*

Among HEDs, olivine has been primarily found in diogenites, commonly associated with orthopyroxene and some accessory minerals like troilite, chromite, silica, iron-nickel (e.g., Mittlefehldt 1994; Bowman et al. 1997; Irving et al. 2009; Beck and McSween 2010; Beck et al. 2011, 2012, 2013; McSween et al. 2011). A survey of olivine-rich HED meteorites shows that there are about 30 diogenites with ≤25 wt.-% olivine (generally less than 10%), 8 diogenites with 40-68% olivine, and 4 dunites with >90% olivine (Floran et al. 1981; Sack et al. 1991; Mittlefehldt 1994; Bowman et al. 1997; Bunch et al. 2006, 2010; Irving et al. 2009; Beck and McSween 2010; Beck et al. 2011, 2012, 2013; McSween et al. 2011). The abundance of olivine in diogenites is typically very heterogeneous, and the estimates could be the result of a sampling bias (e.g., Bowman et al. 1997; Irving et al. 2009; Beck et al. 2011, 2012). Eucrites, which are one of the main components in howardites, normally do not contain olivine (Delaney et al. 1980). It is because of the fact that eucritic components have been removed from the parental melt upon fractional crystallization of basaltic magmas (Mason 1962; Stolper 1977; Grove and Bence, 1979; Delaney et al. 1980). Mikouchi and Miyamoto (1997) also suggested that eucrites don't contain olivine except the late crystallized Fe-rich olivines. Very rarely, a few eucrites have been reported with fayalitic olivine veinlets (Barrat et al. 2011; Zhang et al. 2011). Olivine has also been found in a few howardites, in most cases at the level of less than a few percent (Delaney et al. 1980; Beck et al. 2011, 2012, 2014; Lunning et al. 2014). Olivine-rich impact melts in the range ~0-30 vol.-% (containing ~50-75 vol.-% olivine in the melt) are observed in PCA 02 and GRO 95574 howardites (Beck et al. 2011, 2014).

Prior to the arrival of the Dawn spacecraft at Vesta, various attempts were made to detect olivine. McFadden et al. (1977) suggested either the presence of little olivine (<10%) or no olivine on the Vestan surface from ground-based spectra taken in the wavelength range 0.5-1.06 µm. Their conclusion was based on the apparent symmetry of 1 µm absorption feature. Larson and Fink (1975) and Feierberg et al. (1980) didn't find any spectral indication of olivine in their near-infrared spectra. Gaffey (1997) reported an olivine-bearing unit based on rotationally resolved ground based spectra. Binzel et al. (1997) also suggested olivine-bearing regions based on the observations of four-band spectra (0.43-1.04 µm) from Hubble Space Telescope (HST). Shestopalov et al. (2010) predicted up to 6.8% olivine by simulating the spectra from Binzel et al. (1997) and examining the available ground based spectra. However, Li et al. (2010) and Reddy et al. (2010) didn't confirm olivine on Vesta from their four-band HST spectra and ground based near-infrared spectra, respectively. More recently, Le Corre et al. (2013) and Reddy et al. (2013) revisited sites reported to be olivine rich from ground-based and HST observations using data from Dawn FC and VIR instruments. Le Corre et al. (2013) concluded that the olivine-rich unit reported by Gaffey (1997) called Leslie formation corresponds to the ejecta around the crater Oppia. While the observations of this feature from Dawn instruments and Gaffey (1997) agree well with each other, data from Dawn suggests that the feature is likely to be of impact melt origin rather than olivine (Le Corre et al. 2013; Reddy et al. 2013).

*1.2. Olivine versus High-Ca pyroxene in 1 & 2 μm:*

Pyroxene and olivine are common rock-forming minerals of mafic/ultramafic terrestrial bodies. The presence of olivine in any asteroid or meteorite can signify its origin and evolution tracing the igneous or nebular history (Sunshine et al. 2007), while high-calcium pyroxenes (HCPs) can be used to trace the degree of melting and differentiation of the body (Sunshine et al. 2004). The visible and near-infrared wavelength region (0.4-2.5 μm) is widely applied to detect and analyze these mineralogical compositions. In general, olivine spectra show a composite of three overlapping absorption bands in the 1 μm region. It is basically due to $Fe^{2+}$ in which the major broad absorption feature (around 1.05 μm) is attributed to the M2 cation site in the crystal structure, while the weaker absorption features (around 0.85 and 1.25 μm) are attributed to the M1 cation site (Adams 1974; Burns 1970, 1993; Singer 1981; Cloutis et al. 1986; King and Ridley 1987; Reddy et al. 2011b; Sanchez et al. 2014). The spectral features depend not only on the olivine chemistry, where 1 μm absorption moves to longer wavelength with increasing Fe content, but they also depend on physical parameters of the regolith, like grain size and temperature as well as observational parameters such as phase angle (Adams 1975; Burns 1970, 1993; Cloutis et al. 1986; King and Ridley 1987; Sanchez et al. 2014). Pyroxenes have prominent 1 μm and 2 μm absorption features with varying absorption band centers and band depths. They depend on $Fe^{2+}$-$Ca^{2+}$-$Mg^{2+}$ chemistry and the asymmetry of the cations (crystallographic sites) as well as grain size, temperature and phase angle (Adams 1974, 1975; Burns 1993; Singer 1981; Cloutis and Gaffey 1991; Klima et al. 2007; Schade et al. 2004). The low-Ca pyroxenes have absorption band centers near 0.9 μm and 1.9 μm while high-Ca pyroxenes have band centers near 0.98 μm and 2.15 μm (Pieters 1986). The missing absorption feature of olivine in 2 μm is the key to distinguish olivine from pyroxenes using the band area ratio (BAR) approach (Gaffey 1983; Cloutis et al. 1986; Cloutis and Gaffey 1991). It should be noted that very high-Ca pyroxenes, termed spectral type A by Adams (1974) can have reflectance spectra superficially similar to olivine (Schade et al. 2004), however, such pyroxenes have not been detected in HEDs (e.g., Mayne et al. 2009, 2010; McSween et al. 2011). It is also worth to mention that many authors (Duffard et al. 2005; Moroz et al. 2000; Sanchez et al. 2012) discussed how temperature and grain size could affect mafic silicate reflectance spectra and spectral parameters. They suggested caution in implementing the spectral parameters. Several spectral parameters like the HCP index, forsterite index, fayalite index, and olivine index have been developed and applied to the Martian surface (Poulet et al. 2007; Pelkey et al. 2007; Carrozzo et al. 2012). De Sanctis et al. (2013), Palomba et al. (2012a, b, 2013a, b),

Ruesch et al. (2013, 2014) adapted the Martian spectral parameters to identify potential olivine-rich sites on Vesta using the VIR data.

Distinguishing olivine from high Ca-pyroxene using datasets without having full 1-μm absorption band coverage (e.g., Clementine Ultraviolet/Visible or UVVIS, HST Wide Field Planetary Camera/WFPC, Dawn FC) is rather challenging. It is because of their close spectral similarity in the 1 μm absorption band minima. Clementine UVVIS multispectral data has five bands in the wavelength range between 0.41 and 1 μm. Tompkins and Pieters (1999) and Pieters et al. (2001) suggested olivine-bearing lithologies on the lunar surface using Clementine UVVIS data, however it was difficult to distinguish them from high-Ca pyroxene bearing lithologies. The four band HST/WFPC data covering the wavelength range between 0.43 and 1.04 um were used by Binzel et al. (1997), Shestopalov et al. (2008) and Li et al. (2010) to analyze the likely presence or absence of olivine on Vesta. The Dawn FC is comparable to Clementine/UVVIS and HST/WFPC in terms of their wavelength coverage, but the FC has more spectral bands with better spatial resolution.

## 2. Laboratory-derived Spectra

In this work, spectra in the visible and near-infrared wavelength range have been compiled from available data sets of Reflectance Experiment Laboratory (RELAB) at Brown University/USA (http://www.planetary.brown.edu/relab/), Hyperspectral Optical Sensing for Extraterrestrial Reconnaissance Laboratory (HOSERlab) at University of Winnipeg/Canada (http://psf.uwinnipeg.ca/Home.html), and Unites States Geological Survey (USGS) Spectroscopy Lab (http://speclab.cr.usgs.gov/spectral-lib.html). The compilation includes:

(1) 241 spectra of HEDs (45 eucrite, 13 diogenite, and 17 howardite samples) of various grain sizes and of bulk rock samples from RELAB. A few spectra were excluded in this analysis because of inconsistencies/exceptions observed in the FC spectral range (Appendix-1).

(2) 43 spectra of terrestrial olivine (~$Mg_{90.4}Fe_{9.6}$) - orthopyroxene (~ $Mg_{86.8}Ca_{0.4}Fe_{12.8}$) mixtures (Ol-Opx, 10-90% olivine) at 10% intervals in various grain size ranges (<38, 38-53, 63-90, 90-125 μm) from HOSERLab;

(3) Nearly pure olivines (Ol) having various forsterite contents $Fo_{10-90}$ (<45 μm) from RELAB, $Fo_{11}$-$Fo_{91}$ (<65 μm) from USGS and $Fo_{86.8}$ (<38, 38-53, 63-90, 90-125 μm) from HOSERLab;

(4) 46 spectra of synthetic low/high Ca-pyroxenes (Klima et al. 2011) with various wollastonite contents ($Wo_{2-51}$, <45 μm) from RELAB. Only those HCPs have been considered in our analysis that shows Wo-En-Fs compositions, which are indeed observed in eucrites (Mayne et al. 2009, 2010; McSween et al. 2011). HCPs outside the eucritic compositional range has been discarded (Appendix-2). The selected HCPs (>$Wo_{20}$) are termed HCP/HED in this analysis. All Wo-En-Fs sample spectra of synthetic low/high Ca-pyroxenes (Klima et al. 2011) are termed HCP/CPX ($Wo_{2-51}$).

The laboratory spectra are resampled to FC filter band passes as presented in Sierks et al. (2011). Fig. 1A shows normalized spectra of eucrite (ALHA76005, 25-45 µm), diogenite (EETA79002, 25-45 µm), HCP ($Wo_{45}En_{14}Fs_{41}$, ≤45 µm), Ol ($Fo_{90}$, 38-53 µm) and Ol-Opx (60 wt.-% olivine, 38-53 µm). The filter band passes and center wavelengths of the FC are also presented in this figure. The same spectra resampled to the FC band passes are displayed in Fig. 1B. The figure also visualizes the three band parameters, which are defined as follows:

Band Tilt (BT) = ($R_{0.92\mu m}$ / $R_{0.96\mu m}$)

Mid Ratio (MR) = ($R_{0.75\mu m}$ / $R_{0.83\mu m}$) / ($R_{0.83\mu m}$ / $R_{0.92\mu m}$)

Mid Curvature (MC) = ($R_{0.75\mu m}$ + $R_{0.92\mu m}$) / $R_{0.83\mu m}$

; where R(λ) is the reflectance in the corresponding filter.

## 2.1. Band Tilt (BT)

The BT parameter was found to be well suited to distinguish eucrites from diogenites (Thangjam et al. 2013). Here, we will discuss the relevance of this parameter for distinguishing Ol, Ol-Opx and HCPs. The BT parameter values of the individual samples mentioned in section 2 are plotted along the X-axis in Fig. 2. The values of olivine are in the range 1.02-1.26. The values of HCP/CPX samples lie in the range 0.90-1.61, with the majority (95%) falling in the range 1.01-1.61. The values of HCP/HED samples are in the range 1.11-1.61. Most of the olivine (95%) and HCP samples (89%) have larger BT values than eucrites (0.91-1.03). Eucrites, olivines and HCP samples have higher BT values than diogenites (0.80-0.91). The olivine-orthopyroxene mixtures have BT values in the range 0.75-1.13, while those samples above 40 wt.-% olivine have higher values (0.92-1.13) than diogenites. In this study, we followed the nomenclature of olivine-orthopyroxene mixtures/rock assemblages in accordance with the IUGS system (Streckeisen 1974; Wittke et al. 2011; Mandler and Elkins-Tanton 2013), i.e. olivine-orthopyroxenites (<40%), harzburgites (40-90% olivine), and dunites (>90% olivine). We use the term diogenites to denote the olivine-free orthopyroxenites (Appendix-3). Based on our analyses, the BT parameter is effective in separating peridotites, HCPs and eucrites from diogenites as well as from olivine-orthopyroxenites. HED sample spectra seldom reach a BT value larger than 1.03.

The influence of grain size on the BT parameter is shown in Fig. 3A. The BT values of eucrite ALHA76005, howardite EET875003, diogenite EETA79002 (over the size intervals ≤25, 25-45, 45-75, 75-125, 125-250, 250-500 µm), and Ol-Opx mixtures (over the size intervals ≤38, 38-53, 63-90, 90-125 µm) are presented. The trend of the BT values with increasing grain size of the three HED samples is similar. The BT values of the grain size range 25-45 μm and 125-250 μm are the extremes. There is no significant variation of the values from 125-250 μm to 250-500 μm, while the variations below 125 μm are larger. The BT values of Ol-Opx mixtures do not show a parallel systematic trend. The maximum variations over the whole range of grain sizes are given in brackets (see Fig. 3A). The maximum variation is 7.7% for the three HEDs, 8.2% for Ol-Opx mixtures and 11.7% for pure olivine. The influence of grain size on the BT parameter of HEDs and Ol-Opx mixtures is found to be similar.

*2.2. Mid Ratio (MR)*

The MR parameter values for the individual samples given in section 2 are plotted along the Y-axis in Fig. 2A. Eucrites have MR values between 0.77 and 0.99, while diogenites range between 0.86 and 1.42. Howardites (0.84-1.07) lie between eucrites and diogenites. The values of olivines range between 0.89 and 1.08, while the range is 1.01-1.58 for Ol-Opx mixtures. The values of HCP/CPX samples are in the range 0.44-0.96, while HCP/HED values lie in the range 0.5-0.93. The olivine-orthopyroxene mixtures have larger MR values than HCPs, and therefore these two are distinguishable.

The influence of grain size on the MR parameter is shown in Fig. 3B. The MR parameter of the three HED samples increases with increasing grain size up to 90-125 μm, but decreases thereafter. The values for grain size ranges ≤25 μm and 90-125 μm are the extremes. There is no significant change of the MR values from 125-250 μm to 250-500 μm. A systematic trend is not noticeable for Ol-Opx mixtures (see Fig. 3B). The MR parameter of Ol-Opx mixtures increases with increasing grain size (up to 63-90 μm), while the values are increasing or decreasing for the size range 90-125 μm. The overall influence of grain size on MR parameter decreases with increasing olivine content. The effect is the least (3.8%) for pure olivine, whereas the maximum effect (24.7%) is observed for 10 wt.-% olivine Ol-Opx mixture. Among HEDs, the effect increases from the eucrite sample (8.0%) to the diogenite sample (14.2%).

*2.3. Mid Curvature (MC)*

The MC parameter values for the samples given in section 2 are plotted along the Y-axis in Fig. 2B. Eucrites have MC values in the range 1.81-2.12 while the values of diogenites range between 1.94 and 2.67. Howardites have an intermediate range (1.88-2.30) between eucrites and diogenites. The MC values of olivine are in the range 1.89-2.11 while the range of olivine-orthopyroxene mixtures is 2.03-3.21. The values of HCP/CPX are between 1.57 and 2.54, while the values of HCP/HED lie in the range 1.64-1.94. The majority of eucrites (83%) and HCP/CPX (87%) have lower MC values than Ol-Opx mixtures. However, HCP/HED is distinguishable from Ol-Opx mixtures.

The influence of grain size on the MC parameter is shown in Fig. 3C. Although the values of the three HED samples vary over the whole grain size ranges, there is no significant change from 125-250 μm to 250-500 μm. The MC values of Ol-Opx mixtures increase with increasing grain size (from ≤38 µm to 63-90 µm), but the trend for the grain size range 90-125 µm are different. The overall influence of grain size on the MC parameter decreases with increasing olivine content. The effect is the least (4.3%) for pure olivine, while it is maximal (23.2%) for 40 wt.-% olivine Ol-Opx mixture. Among HEDs, the effect increases from the eucrite sample (5.9%) to the diogenite sample (10.4%).

*2.4. Band depth and albedo at 0.75 µm*

The optical parameters, band depth and albedo, can characterize the abundance of mafic minerals (Pieters et al. 2001). However, these parameters are highly affected by grain size, temperature, viewing geometry and the presence of opaques (Nathues 2000; Reddy et al. 2012a; Hiroi et al. 1994; Duffard et al. 2005; Thangjam et al. 2013). The ratio of the reflectance values at 0.75 and 0.96 (or 1.0) µm is used as a proxy to 1-µm absorption band depth for datasets like lunar Clementine/UVVIS (Tompkins and Pieters 1999; Pieters et al. 2001; Isaacsson and Pieters 2009). In the present work, the ratio of reflectance values of the filters 0.75 and 0.92 µm is used to define the apparent band depth. Figure 4 shows spectra of olivine-orthopyroxene mixtures at 10 wt.-% intervals for the grain size ranges ≤38 and 90-125 µm. For grain sizes ≤38 µm, the band depth and reflectance at 0.75 µm behaves linearly, i.e. with increasing olivine, the albedo gradually increases while the band depth decreases (Fig. 4A, B). For larger grain sizes (90-125 µm), the trend is less systematic (Fig. 4C, D). The influence of grain size on band depth and albedo for the three HED samples (≤25, 25-45, 45-75, 75-125, 125-250, 250-500 µm) and Ol-Opx mixtures (≤38, 38-53, 63-90, 90-125 µm) are analyzed similarly to the band parameters discussed above (Figure not shown). The band depth values vary up to 23.4% for the HED samples and 37% for Ol-Opx mixtures, while the albedo values vary up to 23.3% for the HED samples and 63.4% for Ol-Opx mixtures. Figure 5 shows a scatter plot of band depth and reflectance values for all the samples described in section 2. It is obvious that band depth versus reflectance at 0.75 µm is not suited to distinguish the samples in this analysis.

*2.5. Band parameter approach*

Based on the laboratory data analysis, we conclude:

(1) The BT parameter is effective in distinguishing eucrites from olivine-rich Ol-Opx mixtures (harzburgites), HCPs and eucrites from diogenites and olivine-poor Ol-Opx mixtures (olivine-orthopyroxenites). The dunites (>90% olivine) and HCPs have often larger BT values than eucrites. For BT values above 1.03, dunites and HCPs are obviously separated from HEDs.

(2) The MR values of Ol-Opx mixtures are larger than HCP/HED and HCP/CPX, which mean that the MR parameter is suited to distinguish them from HCPs.

(3) The majority of eucrites (83%) and HCP/CPX (87%) have lower MC values than Ol-Opx mixtures, while HCP/HED samples have lower values than Ol-Opx mixtures.

The combination of the above band parameters is found to be useful to identify olivine-rich Ol-Opx mixtures (peridotites) in the geological context of Vesta (Fig. 2). The band parameter space BT versus MR (BT-MR polygons, Fig. 2A) is more appropriate than BT versus MC (BT-MC polygons, Fig. 2B), because the separation of peridotites from HCP/CPX and HCP/HED are clearer in BT-MR parameter space. Since uncertainties of the laboratory datasets are unavailable, we assumed a 1% standard deviation error. The error propagation for each band parameter is computed statistically (Appendix-4), and the polygons were drawn accordingly (Fig. 2). A quantitative analysis of the influence of grain size on the band parameters for HEDs, Ol, and Ol-Opx mixtures is also presented (Fig. 3). It is worth to mention that the influence of grain size on the BT and MR parameters are lower when compared to the MC parameter. This is one of the reasons why we prefer the use of BT-MR polygons rather than BT-MC polygons. The polygons defined in band parameter space consider various grain sizes and bulk samples (see section 2). Given the fact that the spectra of the laboratory samples used in our analyses are not entirely representing the whole compositional range on Vesta, there may be changes in the polygons defined here.

**3. FC data analyses**

The Dawn Framing Camera (Sierks et al. 2011) acquired images of the entire visible surface of Vesta in three different orbits at spatial resolutions of ~250 m/pixel, ~60 m/pixel, and ~20 m/pixel. There are three standard levels of FC images from which level 1c is processed correcting the "in-field" stray light component (Kovacs et al. 2013). Level 1c I/F data is used for processing in the Integrated Software for Imagers and Spectrometers/ISIS (Anderson et al. 2004) pipeline, where the photometric corrections of the FC color data are performed to standard viewing geometry using Hapke functions. The resulting reflectance data are then map-projected in various steps, and co-registered aligning the color frames to create the color cubes. For the photometric correction, the Vesta shape model derived from FC clear filter images (Gaskell 2012) is used. Further descriptions of the data processing method and the photometric corrections are presented in Nathues et al. (2014b). The FC mosaics generated by the ISIS pipeline were analyzed using ENVI software. For the present analysis, FC color data having ~60 m/pixel spatial resolution from HAMO and HAMO-2 phase are used.

The global mosaic of Vesta in the Claudia-Coordinate system is shown in Fig. 6 using the cylindrical projection. The approximate outlines of the Rheasilvia and the Veneneia basins are marked. Uncertainties of each individual FC color filters were estimated from homogeneous, small areas of different size (2 x 2 to 7 x 7 pixels) at Bellicia and Arruntia. We observed that the relative statistic error for the 4 x 4 pixel sized area is reliable, and these values are used to compute the error propagation of the band parameters (Appendix-4).

*3.1. Arruntia crater*

Arruntia is an impact crater of ~12 km diameter and 2.5 km depth in the northern hemisphere (Fig. 6). A perspective view of the reflectance image at 0.55 µm, projected on HAMO DTM (~62 m/pixel resolution) is shown in Fig. 7A. Potential olivine-rich exposures are highlighted in red by selecting those pixels that have band parameters in the peridotitic field. A few sites are selected for illustration (A1-A5; Fig. 7A) and their average absolute and normalized reflectance spectra over a region of 2 x 2 pixels are presented in Figs. 8A and 8B, respectively. The Vesta average spectrum from FC HAMO-1 & -2 is also displayed. The band parameter space of the identified sites is illustrated in Fig. 9. Olivine-rich exposures are located on the ejecta blanket nearby the outer rim, and a few of them are located on the inner wall and the crater floor. Many of the exposures extend a few hundreds of meters in length, and the exposure marked A3 extends up to few kilometers. The olivine-rich exposures cover ~1.6 % of the area within 2.5 crater radii from the center of the crater. The exposures have higher reflectance value than the average surface of Vesta. In general, the band depth of the olivine exposures is similar to that of the average Vesta, and sometimes slightly deeper. However, a few sites exhibit shallower band depth than the average Vesta. The exposures exhibit a redder visible slope compared to the average Vesta spectrum, which could be due to the associated lithological background materials in the regolith. Dark material is observed in the ejecta blanket nearby the crater rim, and on the slopes of inner crater wall. The lithological background of the olivine-rich exposures is investigated in the band parameter space. The band parameter space of the olivine-rich exposure A3 (located in the ejecta nearby the outer rim, Fig. 10A, B) is presented using the BT-MR polygons (Fig. 10C). The Arruntia region is howarditic/eucritic in composition.

*3.2. Bellicia crater*

Bellicia is an impact crater located westward of Arruntia, having a diameter of ~35 km and 5.9 km depth (Fig. 6). A perspective view of the reflectance image at 0.55 µm is displayed in Fig. 7B. The potential olivine-rich exposures are marked in red. Some sites have been selected (B1-B5, Fig. 7B), and their average absolute and normalized reflectance spectra over a region of 2 x 2 pixels are presented in Figs. 8C and 8D, respectively. The average Vesta spectrum is also shown. Many of the olivine-rich exposures are located on the slopes of the inner crater wall, and a few of them are possibly on the crater floor and in nearby small fresh craters. The exposures extend few hundreds of meters, while some of the exposures (e.g., B1 and B4) are up to few kilometers. The exposures cover ~0.7 % of the crater area within 1 crater radius from the center of the crater. The exposures in Bellicia exhibit higher reflectance values than the average Vesta surface. The exposures in general have similar to slightly higher band depth than the average Vesta. Dark material nearby the olivine exposure B1 are apparently moving along the slope of the inner crater wall (Fig. 7B). The exposures exhibit a redder visible slope compared to the average Vesta spectrum, while two of the sites (B1 and B2) have both higher reflectance values and redder visible slopes than the rest of the exposures (Fig. 8C, D). The locations of the data points in the band parameter space (BT-MR polygons, and BT-MC polygons) are displayed in Fig. 9. The background materials of olivine-rich exposures are analyzed in the band parameter space. The band parameter space of the olivine-rich site B1 (located along the slope of the inner crater wall, Fig. 11A, B) is presented using the BT-MR polygons (Fig. 11C). The majority of the data points are in the howarditic/eucritic field.

## 4. Discussion

Our analysis using FC data suggests olivine-rich exposures at Bellicia and Arruntia. The exposures have higher reflectance values, and similar or slightly higher/lower 1-µm band depths than the average Vesta spectrum. The exposures at Arruntia have redder visible slope than that of Bellicia, which is likely due to the background lithology. The red slope of the ejecta materials at Arruntia could also be caused by an association of impact melt component (Le Corre et al. 2013). The spatial extent of the olivine-rich exposures is found in the range of a few hundred meters up to few kilometers. The exposures are located on the inner crater walls, on the floor of Arruntia, in the ejecta, and in nearby fresh small impact craters. It is to be noted that smaller planetary bodies reveal deep seated minerals, like olivine and spinel, on inner crater walls, central peaks, crater floors, ejecta, and in the vicinity of basins (e.g., Koeppen and Hamilton 2008; Pieters et al. 2011; Yamamoto et al. 2010, 2012). A quantitative analysis of mineralogy and olivine abundance of the exposures seems difficult using FC color data. Moreover, the influence of other factors like grain size has to be considered. However, based on the locations of the data points over the band parameter space (Fig. 9), the identified sites suggest peridotitic lithologies with modal olivine contents above 60%. Such an olivine content is in accordance with the predicted abundance of olivine (60-80%) in the mantle material of Vesta by Mandler and Elkins-Tanton (2013). The exposures at Bellicia and Arruntia could be potential mantle material, but the abundance of olivine is not a sufficient criterion for a mantle origin. The exposures are in general associated with a howarditic/eucritic environment.

*4.1. Source of the olivine-rich material:*

*(1) Excavations from a nearby old impact basin*

The peridotitic exposures at Bellicia and Arruntia are close to the rim of an old basin identified by Marchi et al. (2012). They interpreted this old basin as one of the largest impact structures in the northern hemisphere. It extends ~180 km across with a relative depth of ~10-15 km. They also suggested another large basin nearby the old basin. Assuming an excavation depth in the order of 10-15 km (Marchi et al. 2012) and a crustal thickness of 15-20 km (McSween et al. 2013a), the olivine-rich materials could be excavates of such a basin followed by recent impacts. Meanwhile, olivine is almost absent in the huge Rheasilvia basin (Nathues et al. 2014a; Ruesch et al. 2014), and it seems to be unlikely that even smaller impacts excavated mantle materials. However, Cheek and Sunshine (2014) suggested that the olivine-rich exposures at Bellicia and Arruntia support shallow crustal origin, probably signifying a late stage serial magmatism. The idea of crustal thickness or density variations as well as the petrogenetic model of serial magmatism on Vesta is strengthened by the recent observations of Dawn geophysical data (Raymond et al. 2014b). On the other hand, De Sanctis et al. (2014) suggested that the olivine-rich exposures at Bellicia's crater wall are hard to explain in terms of crustal pluton origin.

*(2) Exogenic origin*

The olivine-rich exposures in the northern hemisphere could be of exogenic origin, delivered by an olivine-rich impactor. The survival of impactor remnants particularly in oblique impacts coupled with low velocity is possible (Bland et al. 2008; Pierazzo and Melosh 2000; Yue et al. 2013). However, the olivine-rich asteroids (A-type) are rare in the main belt (Burbine et al. 1996; Bus and Binzel 2002; Reddy et al. 2011b; Sanchez et al. 2014). The 'missing mantle problem' in the main belt, i.e. the scarcity of asteroidal bodies having composition similar to mantle materials of differentiated and disrupted bodies (Burbine et al. 1996; Sanchez et al. 2014) is a well-known dilemma. Meanwhile, olivine could be an endogenic lithological component of Vesta, which is believed to be abundant in the mantle, as hypothesized from the geochemical/petrological evolution model (e.g., Mandler and Elkins-Tanton 2013; Ruzicka et al. 1997; Righter and Drake 1997; Sack et al. 1991), and the geophysical and thermal models (e.g., Fu et al. 2012; Gupta and Sahijpal 2010; Formisano et al. 2012). Moreover, olivine is comparatively more susceptible to weathering and alteration than pyroxenes when exposed to solar winds and micrometeorite bombardments (Duffard et al. 2005; Yamada et al. 1999). In accordance with these observations, Ammannito et al. (2013a) argued that the exogenic origin of the olivine-rich exposures is unlikely. However, the exogenous origin of olivine cannot be ruled out, and will be discussed in an upcoming paper.

*4.2. Consequences on the geology of Vesta*

Ammannito et al. (2013a) summarized that the finding of olivine-rich exposures at Bellicia and Arruntia in the northern hemisphere and the absence of such materials in the huge basins in the southern hemisphere suggest a more complex evolutionary history compared to pre-Dawn models. Recently, Nathues et al. (2014a) and Ruesch et al. (2014) presented their preliminary observation and mapping of the global distribution of olivine-rich exposures on Vesta using higher spatial resolution FC and VIR HAMO data, respectively. Many of their identified olivine sites are in the northern hemisphere, and only a few are in the Rheasilvia basin. The lack of olivine in the Rheasilvia basin is in contrary to what was expected. However, Beck et al. (2013) suggested that an upper limit of olivine abundance of 30% can be expected on the surface of Vesta within ~60 m exposures (6-16% olivine abundance as more realistic). Furthermore, olivine abundance in diogenites is quite heterogeneous ranging from ~0 to >90 vol.-% (e.g., Bowman et al. 1997; Beck and McSween 2010; Tkalcec et al. 2013). Assuming the olivine abundance suggested by Beck et al. (2013) and the heterogeneity of the distribution of olivine in the regolith of Vesta, it might be possible that the FC and the VIR instruments are not able to spectrally detect them (e.g., Beck et al. 2013; Jutzi et al. 2013).

Meanwhile, Mandler and Elkins-Tanton (2013) argued against the excavation of olivine-rich mantle materials by the Rheasilvia impact. They suggested that the excavation depth in the order of 40 km will excavate all the HED lithologies without the olivine-rich mantle materials assuming their model's crustal thickness of 30-41 km. The non-excavation of olivine-rich mantle materials during the Rheasilvia (and Veneneia impact) was also opined by Jutzi et al. (2013) as an alternative reason to explain the lack of olivine in Rheasilvia. They suggested that Vesta might have a thicker eucritic crust (~100 km) with ultramafic (diogenitic) inclusions.

Contrarily, the chondritic model for Vesta's origin and evolution proposed by Toplis et al. (2013) predicts a relatively orthopyroxene-rich mantle, which is further supported by the observations from Dawn (e.g., Prettyman et al. 2013; Yamashito et al. 2013; Park et al. 2014). Fe abundances (Yamashito et al. 2013) and thermal neutron absorptions (Prettyman et al. 2013) in the Rheasilvia basin and its ejecta observed from the Gamma Ray and Neutron Detector (GRaND) indicate that orthopyroxene-rich lithologies are the excavated materials by the Rheasilvia impact. McSween et al. (2013b) also suggested that the Rheasilvia impact is supposed to excavate the mantle materials, and therefore the occurrence of diogenites in this basin floor (observed from FC and VIR) implies that the mantle materials appear to be excavated and mixed in the ejecta blanket extending across almost half the Vestan surface. McSween et al. (2014) further predicted an olivine-free upper mantle of Vesta because of the lack of spectrally detectable olivine in the Rheasilvia basin.

The evolution of Vesta by serial magmatism (shallow magmatic plutons) is fostered to explain the olivine-rich exposures in the northern hemisphere (Ammannito et al. 2013a; Cheek and Sunshine 2014; Ruesch et al. 2014). The thermo-chemical evolution model of Neumann et al. (2014) predicted the possibility of a shallow magma ocean on Vesta. Again, Barrat and Yamaguchi (2014) argue that the recent magma ocean model proposed by Mandler and Elklins-Tanton (2013) fails to explain the diversity of trace elements observed in diogenites. They suggested that the most likely explanation for the diversity in trace elements in diogenites is by multiple parental melts on Vesta, but not by a magma ocean model. Ruesch et al. (2014) proposed a local enrichment of olivine on Vesta to explain some of the unusual distributions of olivine-rich exposures in the northern hemisphere. De Sanctis et al. (2014) noted that the apparent absence of olivine in Rheasilvia is probably due to the heterogeneity of the Vestan crustal-mantle depths. The observations from Dawn geophysical data reveal significant gravity anomalies that may reflect crustal thickness and/or density variations on Vesta (Raymond et al., 2014b). The observed gravity anomaly and the density variations strengthen the idea of the heterogeneity in the primordial crust and mantle of Vesta favoring multiple plutons within the deep crust or upper mantle (Raymond et al. 2014a, b; Park et al. 2014). Raymond et al. (2014a, b) also suggested that the olivine-rich exposures of Arruntia and Bellicia are part of a northward extension of a strong positive anomaly observed on the eastern equatorial troughs.

Despite all these complexities, the finding of solid-state plastic deformation in olivine grains of diogenite NWA 5480 (57% olivine) and diogenite NWA 5784 (92% olivine) show that they are likely formed in the mantle of Vesta (Tkalcec et al. 2013; Tkalcec and Brenker 2014). Again, Lunning et al. (2014) claimed that the Mg-rich olivine grains found in paired GRO 95 howardites are also of potential Vesta mantle origin. Such findings strengthen the idea of an olivine-rich mantle, which are in accordance with various evolution models (e.g., Ruzicka et al. 1997; Righter and Drake 1997; Sack et al. 1991; Mandler and Elkins-Tanton 2013). The finding of olivine grains in HEDs, which are typical of mantle origin, signifies that the mantle materials were excavated and/or exposed on the surface somehow during Vesta's geological history.

## 5. Conclusion

Dawn FC data can be used to distinguish olivine-rich materials from the howardite-eucrite-diogenite (HED) lithologies on Vesta, despite the FC's limited wavelength range, covering about half of the pyroxene/olivine 1-µm absorption band. Our results are consistent with the recent findings from the VIR data by Ammannito et al. (2013a). The olivine-rich exposures at these localities are mapped using higher spatial resolution Dawn FC color data. The olivine exposures are located on the slopes of the inner/outer crater walls, floor of Arruntia, ejecta, and nearby young small impact craters. The identified sites are potential olivine-rich exposures with modal olivine contents of more than 60%. Both craters are located in howarditic/eucritic background. The recent observation and mapping of globally-distributed potential olivine-rich exposures, including some regions in the Rheasilvia basin (Nathues et al. 2014a; Ruesch et al. 2014) could be a key to better understand the source of the olivine-rich exposures, and consequently the geological evolution of Vesta. However, it remains unsolved why olivine-rich mantle materials were not excavated and/or detected in the Rheasilvia basin as predicted by various models discussed above. Therefore, it calls for in-depth observation and understanding of the composition of the Vestan lithology on a global and local scale from Dawn mission datasets, integrating information from HEDs and the evolution models of Vesta.

**Acknowldegment:**

We acknowledge the entire Dawn mission team, the RELAB and USGS spectral library. E.A.C. thanks the University of Winnipeg, the Canadian Space Agency (CSA), the Canada Foundation for Innovation and the Manitoba Research Innovations Fund for establishment of HOSERLab, and CSA and the Natural Science and Engineering Research Council of Canada (NSERC) for supporting the laboratory spectral data acquisitions and analysis. T.G. is grateful to International Max-Planck Research School, Max-Planck Institute for Solar System Research Institute (IMPRS/MPS). We would like to thank the reviewers (P. Harderson and an anonymous reviewer) and A.E. (Carle Pieters) for their constructive comments and suggestions.

**Appendix:**

1. Among HEDs, few spectra with the RELAB file names CBMS48, CCMS48, CBMS49, CCMS49, C3MS52, L3MS52, L6MS52, C9MS52, L9MS52, CAMS52, LAMS52, C1MS52, L1MS52, C4MS52, L4MS52, C7MS52, L7MS52, CATB20, LATB20, C1TB27, LATB27, MGP023, MGP025, MGP027, MGP029, MGP031, MGP033, MGP035, MGP037, MGP039, MGP041, MGP043, MGP045, MGP047, MGP049, MGP051, MGP053, MGP055, MGP057, MGP059, MGP061, MGP063, MGP065, MGP067, MGP069, CAMP76, CAMP70, CAMP84, CAMP71 are not considered in our analyses. i) Laser irradiated spectra for Millbillillie eucrite (MS-JTW-048-B/CBMS48, MS-JTW-048-C/CCMS48, MS-JTW-052-3/C3MS52, MS-JTW-052-3/L3MS52, MS-JTW-052-6/L6MS52, MS-JTW-052-9/C9MS52, MS-JTW-052-9/L9MS52, MS-JTW-052-0/CAMS52, MS-JTW-052-0/LAMS52, MS-JTW-052-1/C1MS52, MS-JTW-052-1/L1MS52, MS-JTW-052-4/C4MS52, MS-JTW-052-4/L4MS52, MS-JTW-052-7/C7MS52, MS-JTW-052-7/L7MS52) and Johnstown diogenite (MS-JTW-049-B/CBMS49, MS-JTW-049-C/CCMS49) are excluded because the impulse laser treatment yields quite different alteration products, and after all, the correlation of such experimental results with the observed spectra of HEDs are poorly understood (Wasson et al. 1997, 1998). On the other hand, Vesta exhibits a distinctive style of space weathering that differs from other airless bodies since there is no evidence for (lunar-like) nano-phase iron on its regolith (Pieters et al. 2012). It is worth mentioning that if we include all these irradiated spectra, the BT-MR and BT-MC polygons don't change significantly (Fig. 2). ii) The Macibini clast 3 melted/quenched spectra TB-RPB-027/C1TB27 and TB-RPB-027/LATB30 are excluded since the spectra look quite unusual and are significantly different from Macibini clast 3 (TB-RPB-027/C1TB27 and TB-RPB-027/LATB27). iv) C1TB07 and LATB07 (TB-RPB-007; Y792510, ≤1000 μm), C1TB15 and LATB15 (TB-RPB-015; ALHA85001, ≤1000 μm) are excluded because of inconsistencies observed in the individual measurements of the same sample. iii) 24 datasets with the index MGP- are excluded because spectra are not wavelength corrected; instead the corresponding wavelength corrected spectra having file names with the index CGP- are used. iv) The HED samples (EET90020 eucrite (MP-TXH-076-A/CAMP76, TB-RPB-020/C1TB20/LATB20, Cachari eucrite (MP-TXH-084-A/CAMP84), Petersburg howardite (MP-TXH-070-A/CAMP70), and ALHA77256 diogenite (MP-TXH-071-A/CAMP71) which are likely weathered/rusted are excluded.

The spectra used in this analysis are EET87503 (CAMB68, LAMB68A, NCMB68A, CBMB68, CCMB68, CDMB68, CEMB68, CFMB68, CGMB68, CHMB68), Kapoeta (C1MP53, LAMP53, CAMP53), GRO95535 (CAMP67), QUE94200 (CAMP69), EET83376 (CAMP73), EET87513 (CAMP74), Binda (CAMP82), Bununu (CAMP83), Frankfort (CAMP85, CGP049), Le Teilleul (CAMP93, CGP051), Y-7308 (CAMP97), Y-790727 (CAMP98), Y-791573 (CAMP99), GRO95574 (BKR1MP125, C1MP125), QUE97001 (BKR1MP126, C1MP126), Pavlovka (CGP047), Petersburg (CGP053, CGP055), EETA79002, (CAMB67, CBMB67, CCMB67, CDMB67,

CEMB67, CFMB67), Y-74013 (CAMB73, CBMB73), Y-75032 (CAMB74, CBMB74), Johnstown (CAMB95, LAMB95A, CBMB95, CGP057, CGP059, CGP061, CAMS49, C1MS51), Ellemeet (BKR1MP112, C1MP112, BKR1MP113, C1MP113), LAP91900 (CAMP77, C1TB18, LATB18), Aioun el Atrouss (CAMP81), Tatahouine (CAMP88, CGP065, CGP067), A-881526 (CAMP95), Roda (CGP063), Shalka (CGP069), GRO95555 (CAMP68), ALHA76005 (CAMB66, CBMB66, CCMB66, CDMB66, CEMB66, CFMB66, C1TB23, LATB23, C1TB24, LATB24), Millbillillie (C1HH03, C1MB69, C2MB69, C3MB69, CAMB69, LAMB69A, CBMB69, CCMB69, CDMB69, CAMS48, C1MS50, C1RK116A, C1RK116F2, C1RK116G, C1RK116I, C1RK116L), Juvinas (C1MB70, C2MB70, CAMB70, CBMB70, CCMB70, CDMB70, CEMB70, CGP035, CGP037), Y-74450 (CAMB71, CBMB71, CCMB71, CDMB71), ALH-78132 (CAMB72, CBMB72, CCMB72), Padvarninkai (CAMB96, CBMB96, CCMB96, CDMB96, CGP025), Stannern (CAMB97, CBMB97, CGP039, CGP041, CGP043), ALH85001 (CDMB99, CWMB99), Moore County (CAMP86, CAMP86), Pasamonte (CAMP87, CGP033), Bereba (CAMP89, CGP023), Bouvante (CAMP90, C1TB28, LATB28, C1TB29, LATB29, BKR1TB118, C1TB118), Jonzac (CAMP91, CGP029), Serra de Mage (CAMP92), A-881819 (CAMP96), Sioux County (CGP027), Haraiya (CGP031), Nobleborough (CGP045), EET87520 (C1MT29), PCA91078 (C1MT31), Y-792510 (CAMT41), Y-792769 (CAMT42), Y-793591 (CAMT43), Y-82082 (CAMT44), Macibini (C1TB27, LATB27), GRO95533 (CAMP66), PCA82501 (C1TB12, LATB12, BKR1MP124, C1MP124), PCA82502 (C1TB21, LATB21, CAMP80), ALHA85001 (C1TB15, LATB15), ALHA81011 (C1TB14, LATB14, BKR1MP122, C1MP122), ALHA81001 (BKR1MP121, C1MP121, BKR1MT030, C1MT30), LEW87004 (C1TB19, LATB19, CAMP79), Y-75011 (C1TB08, LATB08), EETA79005 (CAMP72, C1TB26, LATB26), EETA79006 (BKR1MP123, C1MP123), LEW85303 (CAMP78), EET83251 (C1TB22, LATB22), EET92003 (BKR1MP118, C1MP118), PCA91006 (BKR1MP119, C1MP119), PCA91007 (BKR1MP120, C1MP120, C1TB16, LATB16), EET87542 (CAMP75, C1TB14, LATB14), Y-791186 (C1TB09, LATB09), A-87272 (CAMP94).

2. Out of 46 spectra of synthetic low/high Ca-pyroxenes ($Wo_{2-51}$, <45 μm) from RELAB (Klima et al. 2011), we selected samples that have reasonable compositions in the contexts of geology of Vesta. i) We consider the calcium bearing pyroxenes free of Fe and Mg as unreasonable. Such pyroxenes neither exist in common basaltic and gabbroic rocks on Earth (Deer et al. 1997), nor in HEDs (Mayne et al. 2009; Mittlefehldt et al. 1998, 2012). Besides, the basaltic Vesta surface in general exhibits prominent 1 μm absorption feature that implies ubiquitous presence of mafic mineralogy. Thus, synthetic pyroxenes without Fs- or without En- components ($Wo_2Fs_{98}$, $Wo_9En_{91}$, $Wo_7Fs_{93}$, $Wo_{10}En_{90}$, $Wo_{17}En_{83}$, $Wo_{25}Fs_{75}$, $Wo_{29}Fs_{71}$, $Wo_{30}En_{70}$, $Wo_{46}En_{54}$, $Wo_{35}Fs_{65}$, $Wo_{38}En_{63}$, $Wo_{39}Fs_{61}$, $Wo_{46}En_{54}$, $Wo_{51}Fs_{49}$) are omitted. ii) We follow Sunshine et al. (2004) and Klima et al. (2011) in restricting the high-Calcium clinopyroxene (HCP) compositions to Wo- contents above 20 mol-%. Clinopyroxene compositions below that value are very rare in eucrites and diogenites and probably result from exsolved HCP lamellae from pigeonite hosts (e.g., McSween et al. 2011). This compositional field is marked by the open box in the Figure given below (Fig. A). This Figure (open circles/plus markers) is based on the compilation of pyroxene compositions in eucrites given by Mayne et al. (2009). Comparing the compositional variations of pyroxenes in eucrites with that of the synthetic pyroxenes investigated by Klima et al. (2011) and Sunshine et al. (2004) reveals that a small number of syntheitic pyroxenes plots outside the compositional variation of diogenites and eucrites (indicated by the gray box). iii) All synthetic pyroxenes plotting outside that compositional field ($Wo_{23}En_6Fs_{70}$, $Wo_{39}En_{52}Fs_9$, $Wo_{45}En_{52}Fs_3$, $Wo_{45}En_{46}Fs_9$, $Wo_{49}En_{45}Fs_6$, $Wo_{49}En_{42}Fs_8$, $Wo_{49}En_{43}Fs_8$, and $Wo_{49}En_1Fs_{50}$) are not considered here.

3. The existing nomenclature of diogenites and/or olivine bearing diogenites by different researchers is ambiguous (Sack et al. 1991; Bunch et al. 2010; Beck and Mcsween, 2010; Wittke et al. 2011; Ammannito et al. 2013a; Mandler and Elkins-Tanton 2013). Here, we follow the nomenclature that is in accordance with IUGS system (Streckeisen 1974; Wittke et al. 2011; Mandler and Elkins-Tanton 2013).

4. The error propagations are computed using the statistical formulations given below-

$$\partial|BT| = |BT|\sqrt{\left(\frac{\partial|0.92\mu m|}{|0.92\mu m|}\right)^2 + \left(\frac{\partial|0.96\mu m|}{|0.96\mu m|}\right)^2}$$

$$\partial|MR| = |MR|\sqrt{\left(\frac{\partial|0.75\mu m|}{|0.75\mu m|}\right)^2 + \left(\frac{\partial|0.92\mu m|}{|0.92\mu m|}\right)^2 + 2\left(\frac{\partial|0.83\mu m|}{|0.83\mu m|}\right)^2}$$

$$\partial|MC| = |MC|\sqrt{\left(\frac{\sqrt{(\partial|0.75\mu m|)^2+(\partial|0.92\mu m|)^2}}{|0.75\mu m|+|0.92\mu m|}\right)^2 + \left(\frac{\partial|0.83\mu m|}{|0.83\mu m|}\right)^2}$$

Where, $\partial|\lambda|$ is the standard deviation, and $|\lambda|$ is the value of the band parameters or reflectance at their respective wavelength. The uncertainties of the band parameters for laboratory spectra (BT- 1.4%, MR- 2%, MC- 1.2%) are shown in Figure 2. In the same way, the uncertainty limits of the band parameters for FC spectra are also computed by selecting nearly homogenous areas, based on similar ranges/values of the BT parameter, topography and reflectance. Four such sites are selected, and spectra are collected for various pixel sizes (2 x 2, 3 x 3, 4 x 4, 5 x 5, 6 x 6 and 7 x 7). The average spectra and their standard deviation are used to statistically compute the error propagation for the band parameters. The uncertainty limits for 4 x 4 pixels are selected after examining the trend of the values for consistency. Therefore, we observed 0.79, 0.78, 0.74, 0.81, 0.81, 0.87, 0.84% uncertainties for the seven filters (in ascending center wavelengths of the filters), and then, the uncertainties for the band parameters (BT- 1.19%, MR- 2.02%, MC- 1.02%) are also estimated (Fig. 9).

**Figure Caption:**

**Fig. 1:** Reflectance spectrum of howardite (EET875003, 25-45 μm), eucrite (ALHA76005, 25-45 μm), diogenite (EETA79002, 25-45 μm), high-Ca pyroxene ($Wo_{45}En_{14}Fs_{41}$, ≤45 μm), olivine ($Fo_{90}$, 38-53 μm) and olivine-orthopyroxene mixture (60 wt.% olivine, 38-53 μm) normalized to unity at 0.75 μm. Framing Camera filter band passes and center wavelengths are marked. (B) Spectra are resampled to FC filter band passes. A sketch defines the band parameters Mid Curvature (MC), Mid Ratio (MR) and Band Tilt (BT).

**Fig. 2:** (A) Mid Ratio versus Band Tilt, and (B) Mid Curvature versus Band Tilt for eucrites, diogenites, howardites, Ol, HCP and Ol-Opx mixtures. The HED samples are in various grain sizes/bulk from RELAB. HCP/CPX samples (<45 μm) are synthetic clinopyroxenes with compositional range $Wo_{2-51}$, while HCP/HED are selected samples compatible with the existing HCPs among eucrites. Ol-Opx spectra (38-53, 63-90, 90-125 μm) are from HOSERLab. Ol spectra (terrestrial olivines) are from RELAB (<45 μm, $Fo_{10-90}$), USGS (<65 μm, $Fo_{11-91}$), and HOSERLAB ($Fo_{90}$ in various grain sizes). Ol and HCP/HED are highlighted by filled symbols.

**Fig. 3:** Influence of grain size on (A) BT, (B) MR and (C) MC parameters, for HEDs in size ranges <25, 25-45, 45-75, 75-125, 125-250, 250-500 μm, and olivine-orthopyroxene mixtures (10-90% Olivine) and 100% olivine in size ranges <38, 38-53, 63-90, 90-125 μm. The maximum variations (%) of each sample over the whole grain size ranges are given in brackets. Data points for some of the mixtures are not shown to enhance readability.

**Fig. 4:** (A) Absolute and (B) normalized spectra of Ol-Opx mixtures (≤38 μm). (C) Absolute and (D) normalized spectra of Ol-Opx mixtures (90-125 μm). Spectra of a few mixtures are not shown to enhance readability (B, D).

**Fig. 5:** Band depth (apparent) and reflectance values at 0.75 μm for howardites, eucrites, diogenites, olivines, olivine-orthopyroxene mixtures, and HCP/HED.

**Fig. 6:** HAMO global mosaic of Vesta in the Claudia-Coordinate system at ~60 m/pixel resolution in simple cylindrical projection. Arruntia (A) and Bellicia (B) craters are in the northern hemisphere. The approximate outlines of the Rheasilvia and Veneneia basins are marked in bold and dashed lines, respectively.

**Fig. 7:** Perspective view of reflectance image of (A) Arruntia and (B) Bellicia crater, projected on HAMO DTM. Potential olivine-rich exposures are marked in red.

**Fig. 8:** Spectra of olivine-rich sites as indicated in Fig. 7. (A) Absolute and (B) normalized spectra from sites at Arruntia. (C) Absolute and (D) normalized spectra from Bellicia. Each spectrum is an average of 2 by 2 pixels. The spectrum of the average Vesta surface (black solid line) is also shown.

**Fig. 9:** Location of data points of olivine-rich exposures at Bellicia and Arruntia crater, projected on (A) BT-MR polygons, and (B) BT-MC polygons. The polygons are based on our laboratory spectral analyses (see Fig. 2).

**Fig. 10:** (A) Arruntia crater, and (B) selected region near the olivine-rich exposure A3. (C) Band parameter values of the selected region plotted over the BT-MR polygons.

**Fig. 11:** (A) Bellicia crater, and (B) selected region near the olivine-rich exposure B1. (C) Band parameter values of the selected region plotted over the BT-MR polygons.

**Figure caption in Appendix:**

**Fig. A:** Pyroxene composition in eucrites (open circles/plus), adapted from McSween et al. (2011) and Mayne et al. (2009). Markers in open squares are synthetic pyroxenes (Klima et al. 2011). We restrict high-Calcium pyroxene compositions (HCP/HED) to Wo- contents above 20 mol-%. The gray box indicates those clinopyroxenes considered in our analysis, while open box represents clinopyroxene compositions below 20 mol-% Wo.

**Fig. 1:**

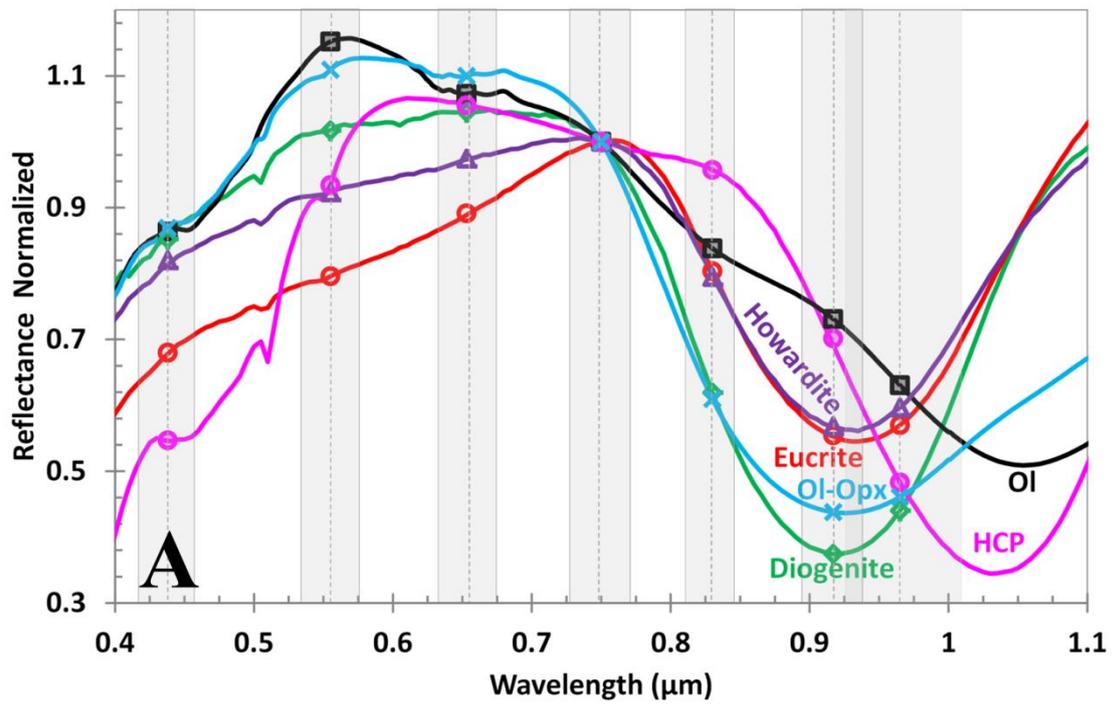

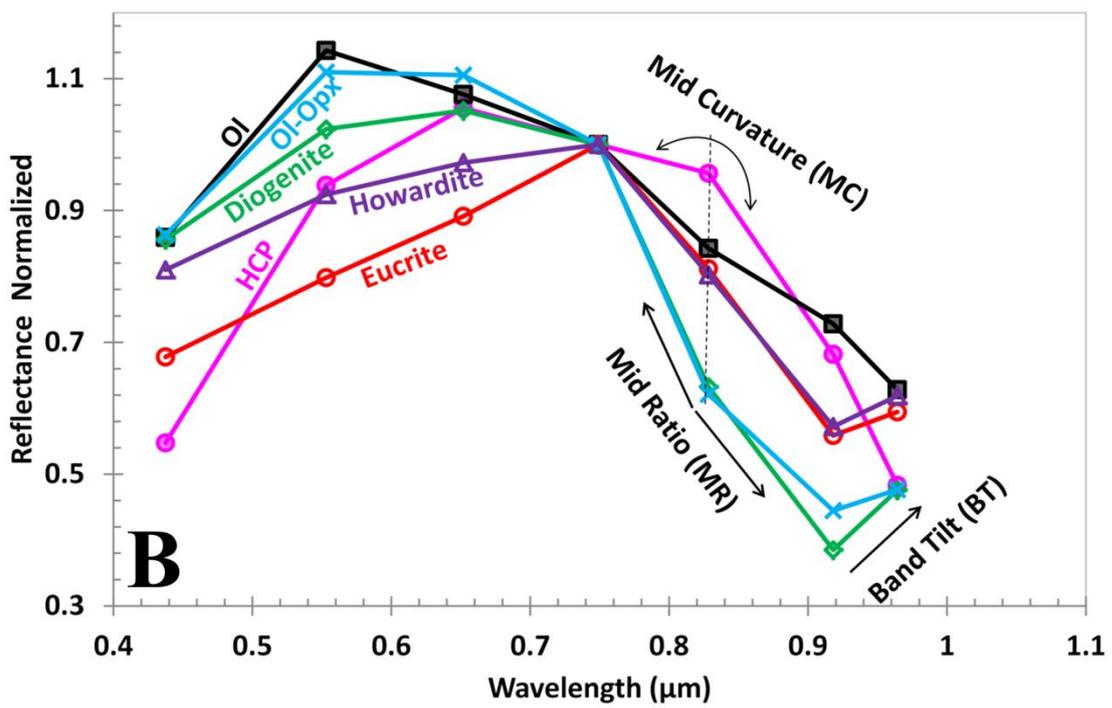

**Fig. 2:**

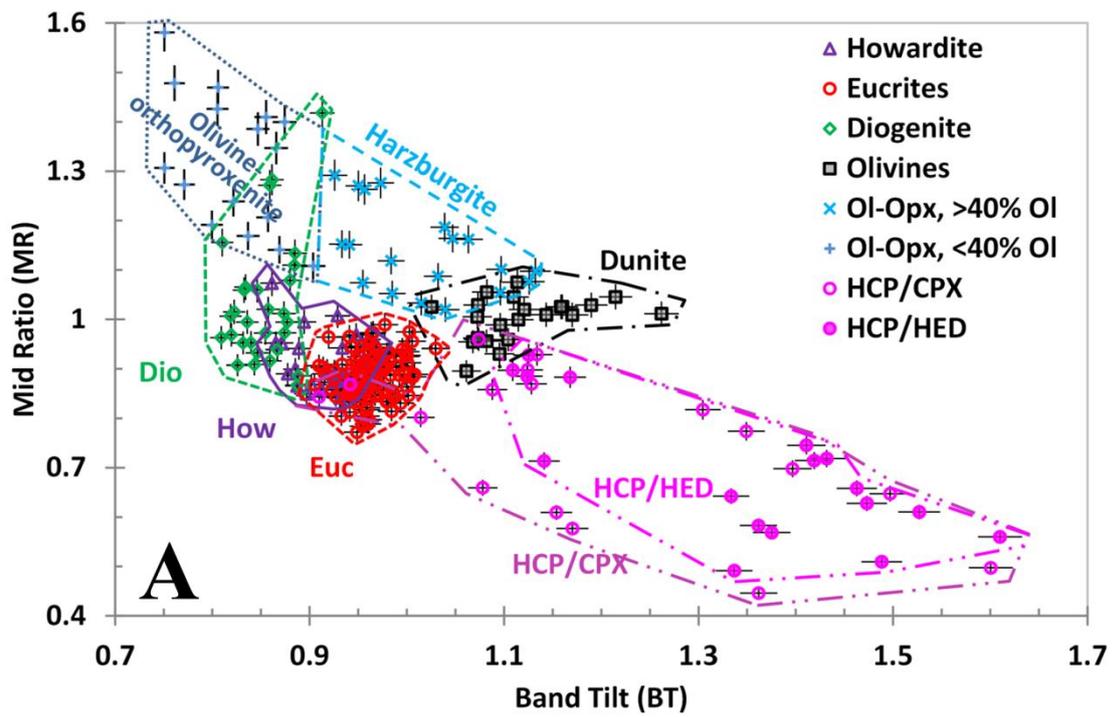

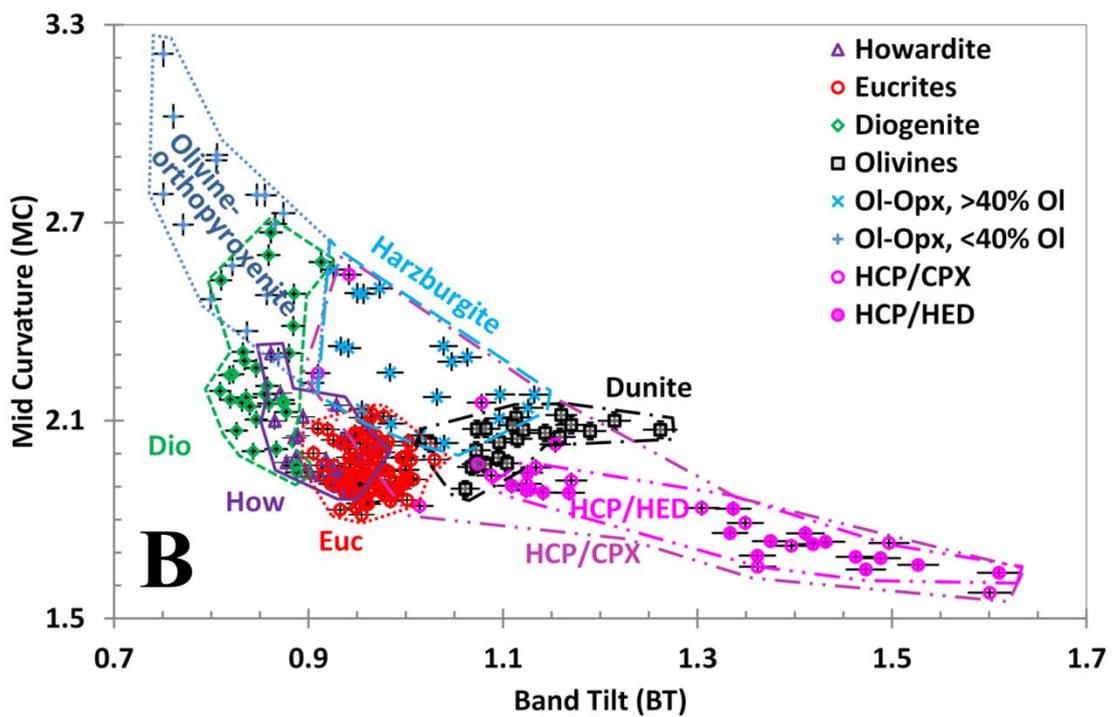

**Fig. 3:**

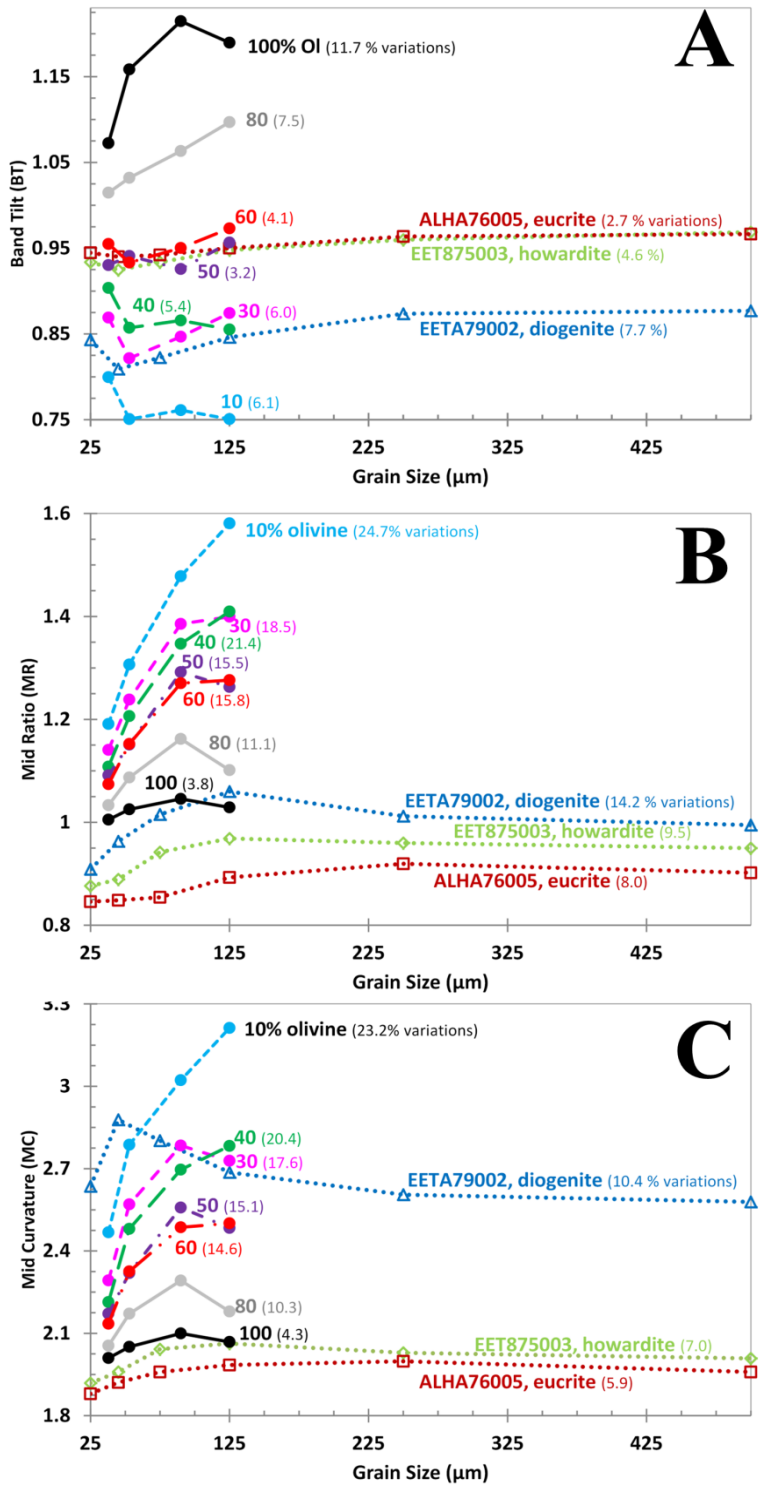

**Fig. 4:**

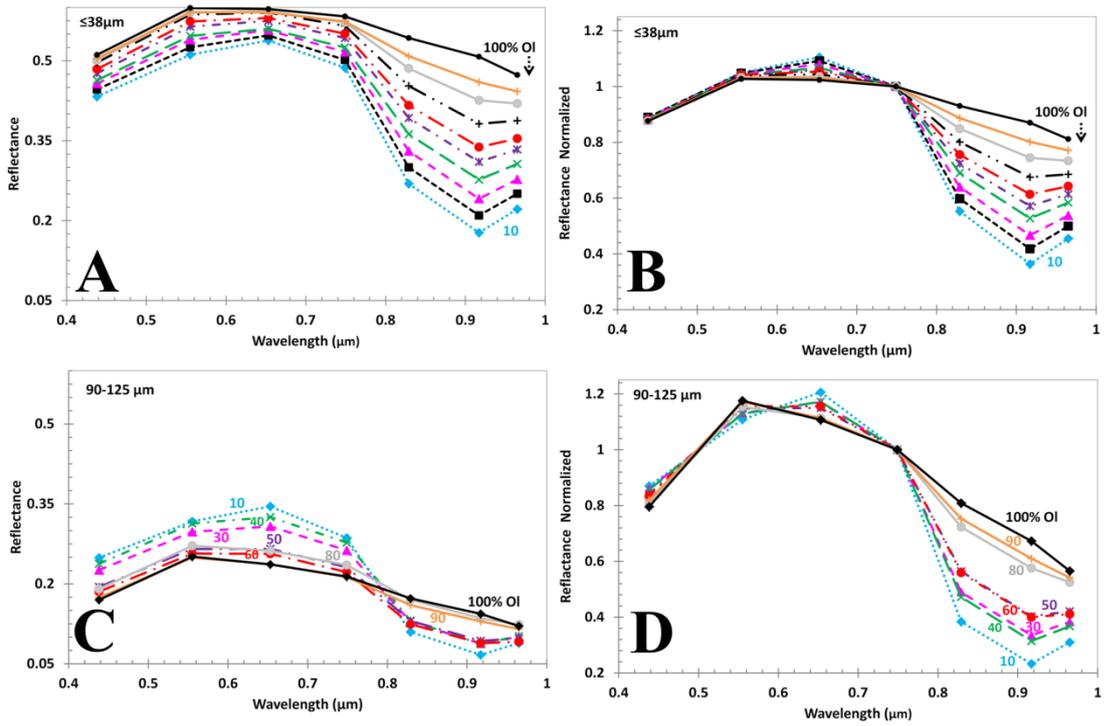

**Fig. 5:**

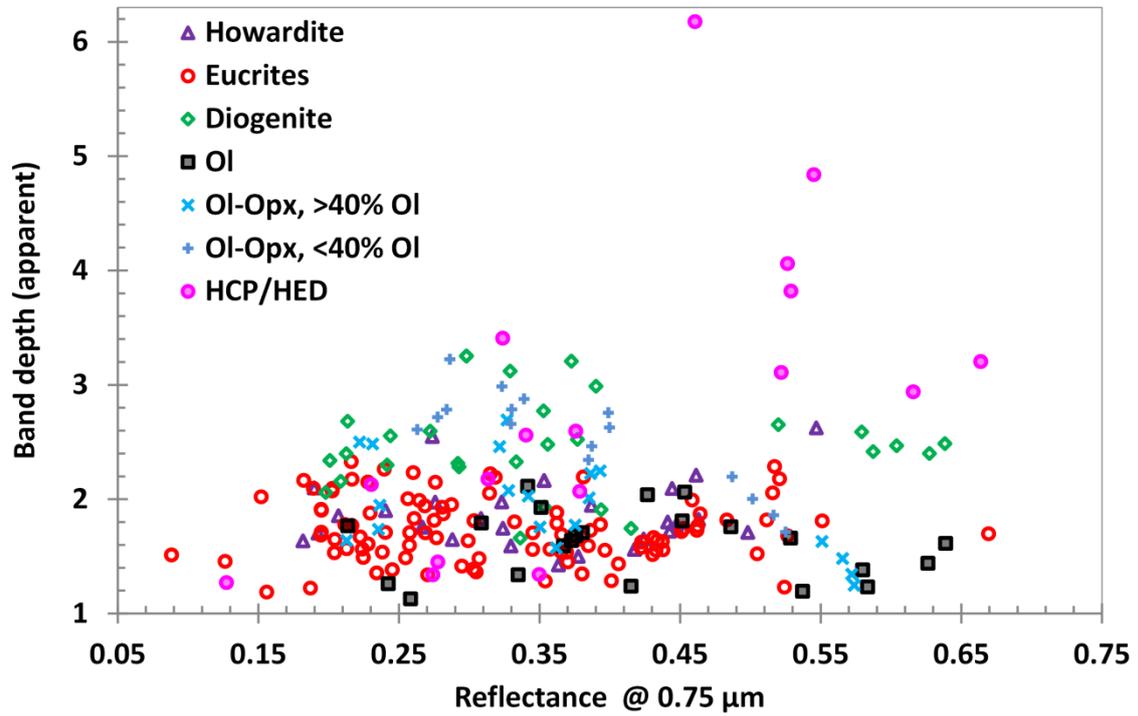

**Fig. 6:**

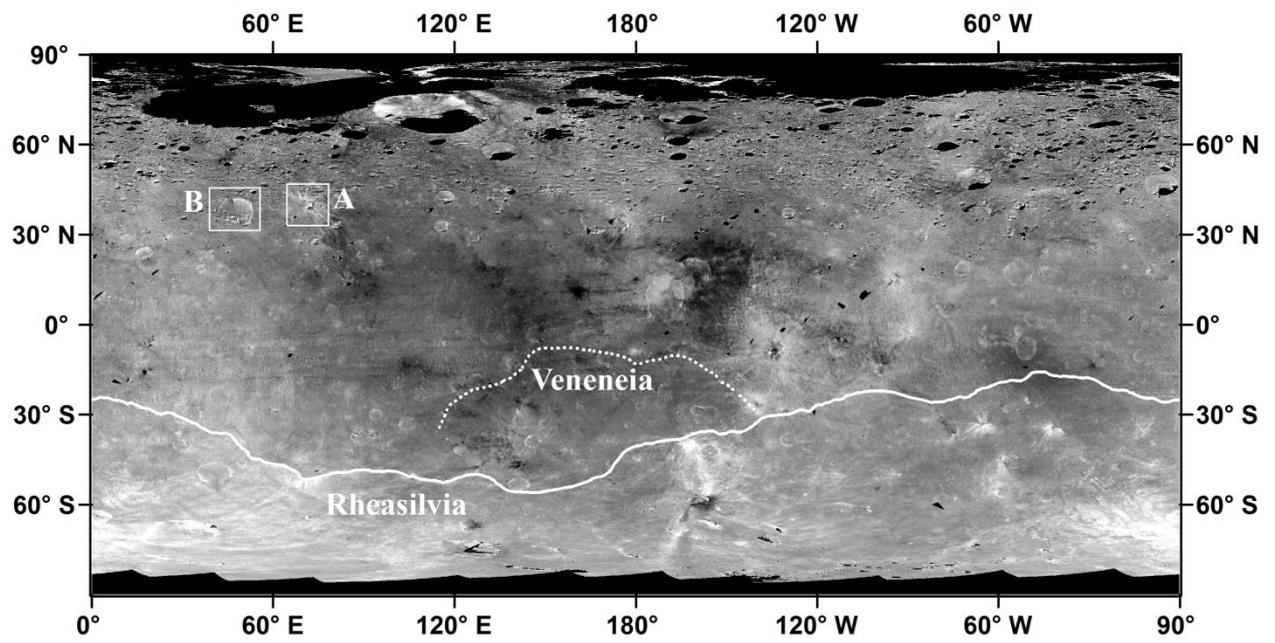

**Fig. 7:**

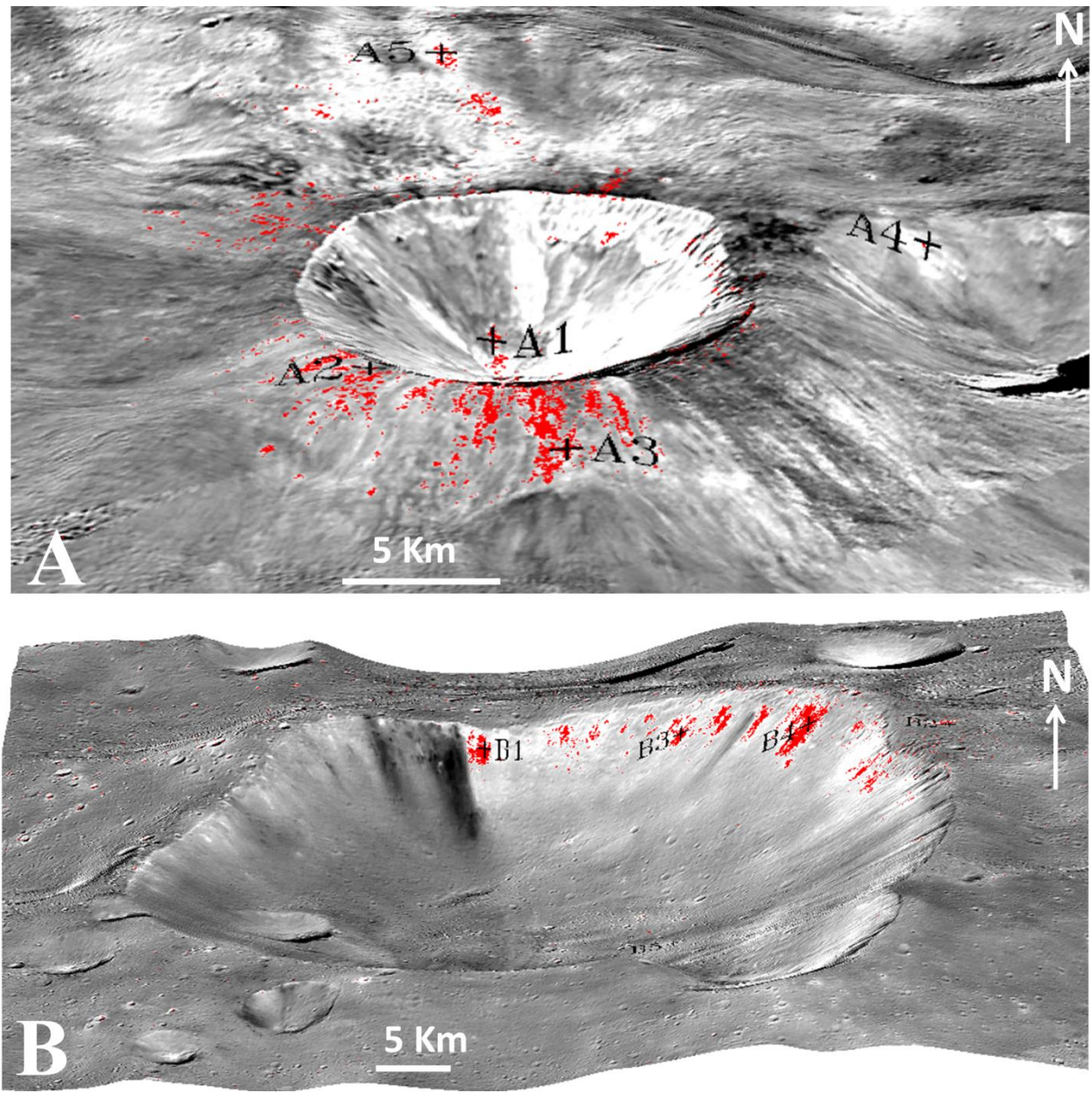

**Fig. 8:**

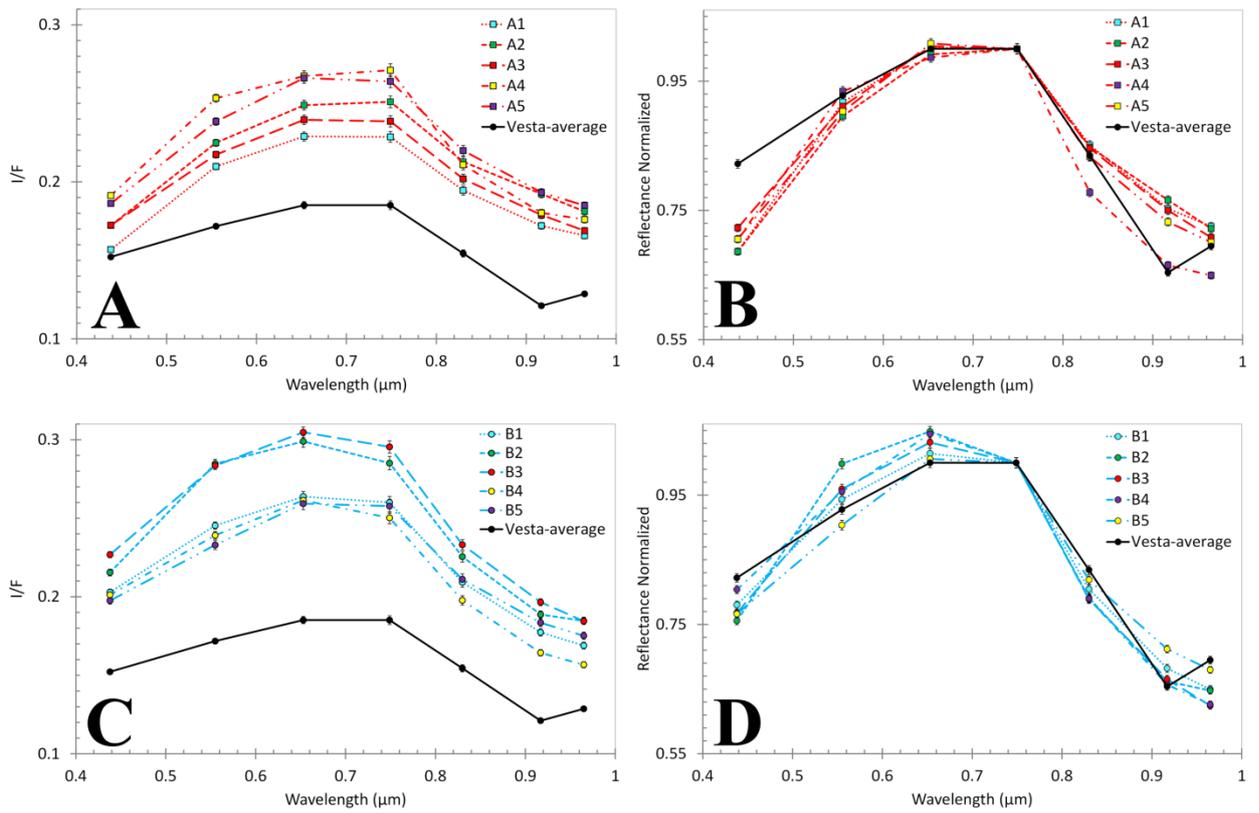

**Fig. 9:**

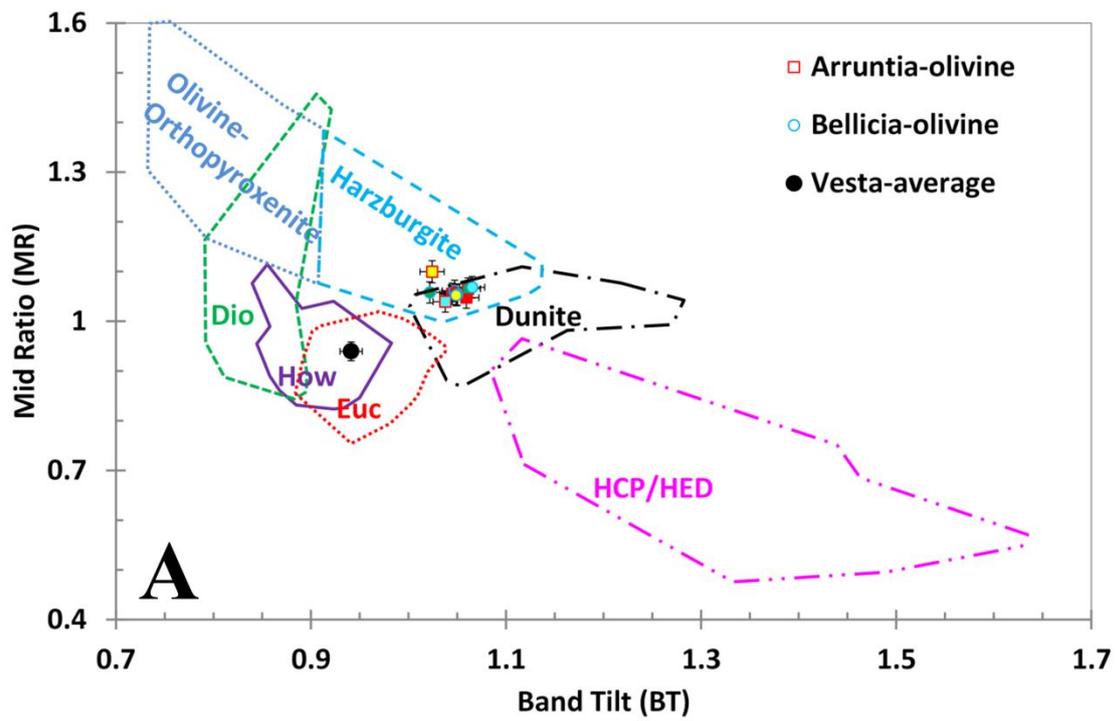

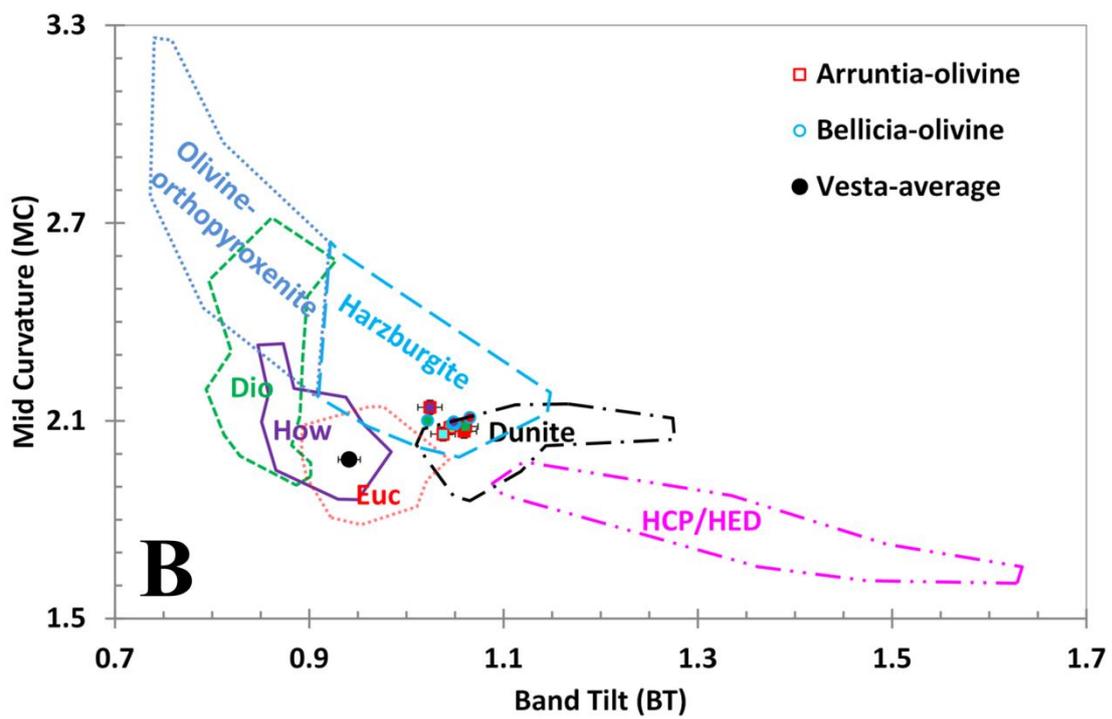

**Fig. 10:**

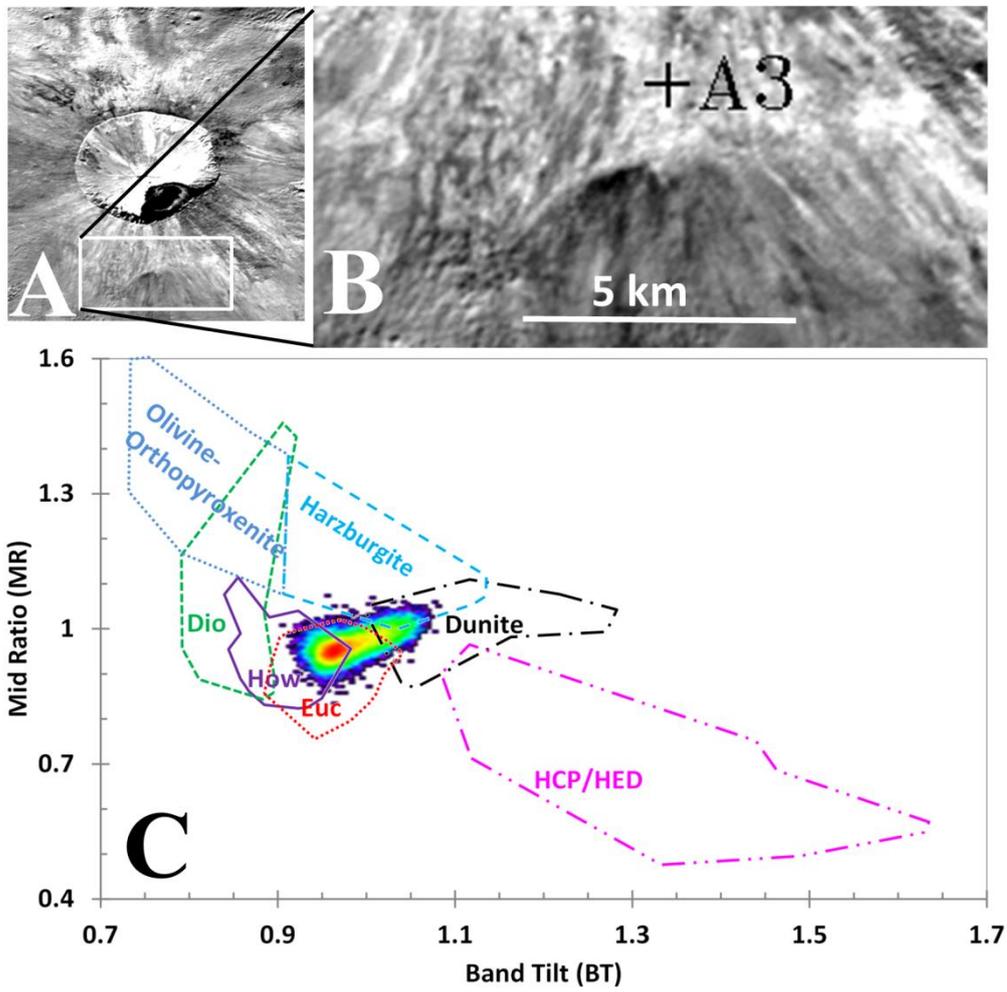

**Fig. 11:**

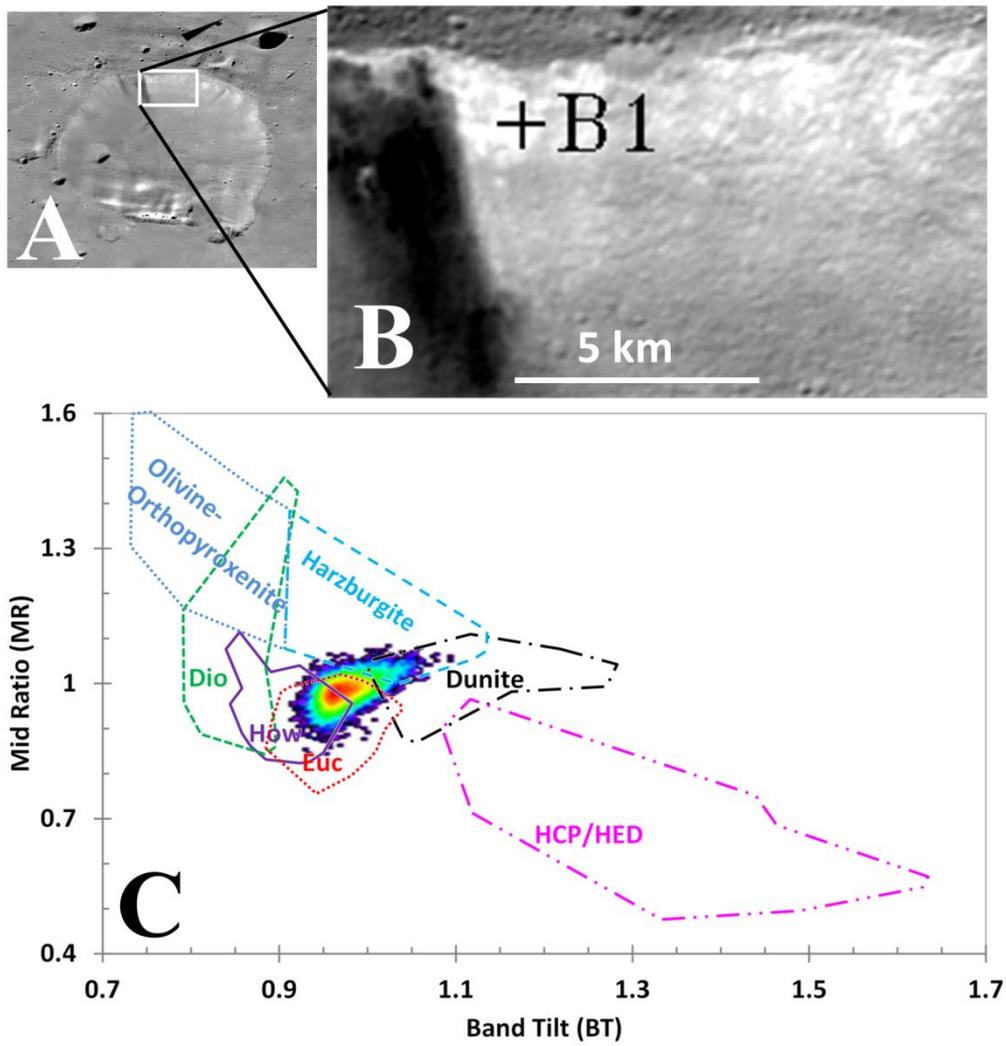

**Fig. A (Appendix):**

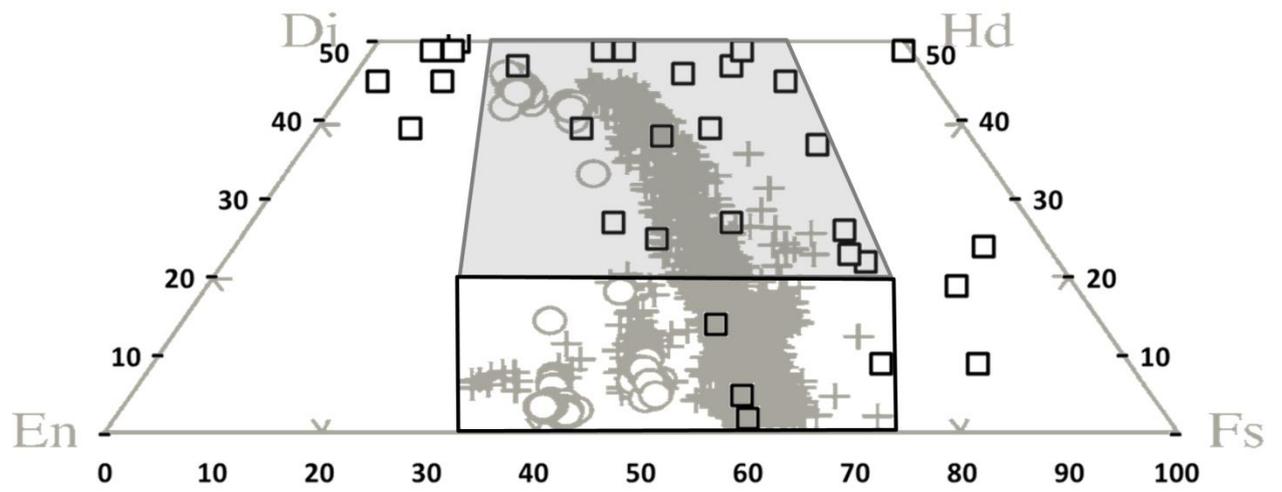